\documentclass[3p]{elsarticle}
\usepackage{graphicx} 
\usepackage{url}
\usepackage{bm}
\usepackage[normalem]{ulem}
\begin{document}

\title{Frequency locking near the gluing bifurcation:\\ 
Spin-torque oscillator under periodic modulation of current}

\author{Michael A. Zaks$^*$}
\ead{mzaks@uni-potsdam.de}
\author{Arkady Pikovsky}
\ead{pikovsky@uni-potsdam.de}
\address{Institute of Physics and Astronomy, D-14476 Potsdam, Germany}

\begin{abstract}
We consider entrainment by periodic force of limit cycles 
which are close to the homoclinic bifurcation. Taking as a physical example
the nanoscale spin-torque oscillator in the LC circuit, we develop the general
description of the situation in which the frequency of the stable periodic orbit
in the autonomous system is highly sensitive to minor variations of the parameter, 
and derive explicit expressions for the strongly deformed borders of the 
resonance regions (Arnold tongues) in the parameter space of the problem. 
It turns out that proximity to homoclinic bifurcations hinders synchronization 
of spin-torque oscillators. 
\end{abstract}
\begin{keyword}
Synchronization \sep homoclinic bifurcation \sep Arnold tongues
\sep non-isochronous oscillators.
\end{keyword}

\maketitle

\noindent $^*$\ Corresponding author. 
Tel.: +49 331 9775479, Fax +49 331 9775947.
\section{Introduction}
Synchronization, commonly present in ensembles of interacting elements, 
is based on the tendency of coupled oscillators to adjust their characteristic 
timescales. Take a system in which several oscillators are synchronized, 
and perform a minor shift of one of the governing parameters. 
If the individual timescale of oscillations  for every element
is only slightly affected by this shift, 
it seems reasonable to expect that the collective state would sustain
synchrony, up to small changes in its quantitative characteristics. 
If, in contrast, the timescale of one or several parts of the ensemble is highly
sensitive to the parameter variation, adjustment to neighbors can get disrupted,
and in the worst case synchronization breaks down.
As a simple example, let every uncoupled unit perform the limit cycle dynamics. 
The situation when each limit cycle is robust and reacts to parameter 
variations only by modest shifts of frequency, 
is well understood (see e.g.\cite{synchro1,synchro2}: 
there, as a rule, synchronization is persistent.
If, on the contrary, some of these limit cycles are fragile: 
stay on the verge of disappearance (via e.g. a saddle-node bifurcation), or if 
their frequencies strongly depend on the parameters (e.g. due to proximity 
to a homoclinic bifurcation), maintaining the synchrony can turn problematic.

In this paper, we take a closer look at the peculiarities of synchronization
in the situation when the oscillating unit is close to the homoclinic
bifurcation. As a physical object, we consider the spin-torque oscillator
(STO) in the LC-circuit under the action of the periodic modulation 
of current.
In recent years spin-torque oscillators have attracted broad interest
as prospective nanoscale generators of  electromagnetic microwaves in the 
gigahertz frequency range.
A spin-torque oscillator is a device composed of two or more
magnetic layers separated by non-magnetic spacers; 
electric current passes across one of such layers, becomes spin-polarized
and exerts torque on the magnetic particles in the other layer, causing
the high-frequency precession of the magnetization vector:  
spin-transfer effect~\cite{Katine_2000}. The effect of 
giant magnetoresistance transforms this precession into microwaves.
Since the field of a single spin-torque oscillator is weak, robust 
synchronous output of many such coupled units is desirable for applications. 
However, bringing  ensembles of spin-torque oscillators to synchrony
proved to be a difficult task.

This seems to be related to the fact that spin-torque nano-oscillators
are highly non-isochronous: the frequency of their oscillations is strongly
dependent on the amplitude~\cite{Finocchio_2012, Quinsat_2012}. 
By introducing an additional variable into 
the conventional Adler equation for the description of phase evolution, 
the authors of ~\cite{Zhou_etal_2010,Slavin_Tiberkevich_2009}
obtained explanations for a number of experimentally observable features, 
including the oscillatory transient dynamics which precedes the onset of
the phase-locked state.
Qualitative and numerical analysis of different configurations of 
the coupled STO gives evidence that in the operative range of relevant
parameters each oscillator is, in a sense, close to homo- and heteroclinic 
bifurcations~\cite{Li_Zhou_2011,Li_Zhou_2012,Pikovsky_2013}. 
Below, we study the basic reasons for the difficulties of robust 
synchronization by considering sort of a minimal model: a single STO 
under the action of the periodic external force. 
It turns out that even the simplest entrainment: that by the periodic
harmonic signal, is strongly hampered by proximity to homoclinics.
Typical symmetry of the STO often ensures that the particular kind 
of homoclinic bifurcation: a so-called ``gluing bifurcation''~\cite{gluing} 
in which stable periodic states coalesce and recombine, takes place.
In the context of serially coupled arrays of STO gluing bifurcations
were reported in~\cite{Turtle_2013}.
Physically, closeness to a homoclinics means that minor changes
of system parameters impel strong variations of characteristic times.
Understandably, these variations hinder synchronization at a fixed frequency. 
Going beyond a particular model, we consider the
general situation in which a periodic force acts upon a dynamical 
system which is close to the gluing bifurcation,
and obtain expressions for the borders of the resonance regions 
(Arnold tongues) in the parameter space. These general
theoretical predictions concern, as a particular case, the
periodically forced STO.

The layout of the paper is as follows.
In Sect.~\ref{sect_sto} we briefly delineate the mechanism which
generates dynamics of the spin-torque oscillators. 
For the solitary spin-torque oscillator in the LC-circuit,
we derive the governing equations and put them into the suitable form.
In Sect.~\ref{sect_unforced} we describe dynamics of that oscillator in
absence of the external modulation. We show that increase of the 
constant current through the unit destabilizes the quiescent state and
replaces it by small-scale oscillations of the magnetic moment.
Further growth of the current leads to the formation of homoclinic
orbits to the saddle equilibrium; in the large range of other parameters 
of the problem, this is a gluing bifurcation. Since in the course
of transition the period of oscillations becomes infinite, entrainment
of these oscillations by an external periodic force 
(which, in our case, is provided by modulations of the current in the circuit)
is not quite trivial.
In Sect.~\ref{sect_mapping} we present a general formalism
for the description of frequency locking near the gluing bifurcation.
To this purpose, we construct the mapping which interrelates
the coordinates of subsequent returns of the orbit into the vicinity
of the equilibrium of the unperturbed system. Analysis of the mapping
discloses the unusual shape of the Arnold tongues
on the parameter plane of the problem. Entrained oscillations
are stable in the region between the saddle-node, period-doubling
and torus (Neimark-Sacker) bifurcations. Under low forcing frequencies,
the distance between the two former bifurcations becomes exponentially
small, and the entrainment, albeit formally present, can hardly be observed. 
The torus bifurcation, in contrast, is a resonant phenomenon, which
occurs close to a certain particular frequency of forcing. 
Further, in the large parts of the parameter space, the entrained dynamics 
is (at least) bistable: two different kinds of periodic oscillations coexist.
To  illustrate these general findings, we return
in Sect.~\ref{sect_forced} to the periodically forced spin-torque oscillator
and numerically determine the borders of Arnold tongues 
in the space of its parameters. 
In the end, along with the general discussion,
we briefly outline the relation of the results to synchronization of
two autonomous coupled STOs.
 
\section{Spin-torque oscillator: governing equations}
\label{sect_sto}
\begin{figure}[h]
\centerline{\includegraphics[width=0.4\textwidth]{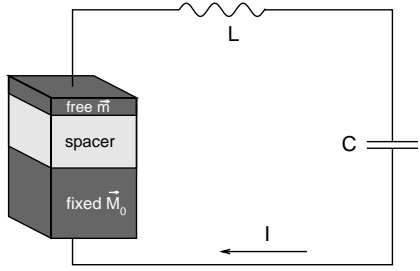}}
\caption{Spin-torque oscillator in the LC-circuit.}
\label{fig_setup}
\end{figure}

Consider a spin-torque oscillator in the LC-circuit (Fig.\ref{fig_setup}). 
In the minimal configuration this is a stack of three layers: two magnetic ones 
and the nonmagnetic spacer between them.
The thicker layer has constant magnetization, 
whereas the magnetization in the thinner one can freely rotate. 
Electric current passes through the thick magnetic layer, and, 
due to the interaction of electrons with magnetic field, 
becomes spin-polarized. 
Then, injected into a thin free magnetic 
layer, this polarized current induces precession of magnetization.
Let us obtain the dynamical equations which govern this process.
Our derivation and notation largely follow \cite{Li_Zhou_2010,Pikovsky_2013}.
Discarding the fine details of micromagnetic interactions inside the free layer,
we restrict ourselves to the macroscopic description in the framework
of the Landau-Lifshitz-Gilbert-Slonczewski magnetization equation.
Within this approach, the unit vector  $\vec{m}$ of magnetization 
in the free layer obeys equation of motion
\begin{equation}
\frac{d\vec{m}}{dt}=-\gamma \vec{m}\times\vec{H}_{\rm eff}
+\alpha\vec{m}\times\frac{d\vec{m}}{dt}
+\gamma\overline{\beta} J\,\vec{m}\times (\vec{m}\times\vec{M}_0)
\label{llgs}
\end{equation}
where the last term, as demonstrated by Slonczewski~\cite{Slonczewski}, is
responsible for the current-driven spin transfer.
Here,  $\gamma$ denotes the giromagnetic ratio, 
$\alpha$ is the Gilbert damping coefficient,
and the effective Landau-Lifshitz field $\vec{H}_{\rm eff}$ is a superposition 
of three components: the external magnetic field $\vec{H}_a$, 
the uniaxial anisotropy field $\vec{H}_k$ directed along the axis of
easy magnetization, and the demagnetizing contribution $\vec{H}_{dz}$.
In the spin-transfer term, 
$\vec{M}_0$ is the constant magnetization of the thick fixed layer, 
$\overline{\beta}$ characterizes the material properties of the free layer, 
and finally,  $J$ is the instantaneous current through the circuit.
Spin-transfer changes back and forth the magnetoresistance of the STO 
(see below),
therefore $J=J(t)$ is, in general, time-dependent, and Eq.(\ref{llgs}), 
taken alone, is essentially non-autonomous.
We align the $x$- and $z$-axes along the directions of, respectively, 
the external field and the demagnetization field. 
Then, in Cartesian coordinates, Eq. (\ref{llgs}) becomes
\begin{eqnarray}
 \mu\,\frac{d m_x}{dt}&=& H_{dz} m_y m_z
       +\alpha\left( (H_a+H_k m_x)(m_y^2+m_z^2)+H_{dz}m_x m_z^2)\right)
       -\overline{\beta} J M_0 (m_y^2+m_z^2)\nonumber\\
 \mu\,\frac{d m_y}{dt}&=& 
    H_{dz}m_z (\alpha m_y m_z -m_x) -H_k m_x (\alpha m_x m_y +m_z)\nonumber\\
   & & -H_a (\alpha m_x m_y+m_z)+\overline{\beta} J M_0 (m_x m_y-\alpha m_z) \\
\mu\,\frac{d m_z}{dt}&=& (H_a+H_k m_x)(m_y-\alpha m_x m_z)
            -\alpha H_{dz}(m_x^2+m_y^2) m_z 
	    + \overline{\beta} J M_0 (m_x m_z+\alpha m_y)\nonumber
\label{cartesian}
\end{eqnarray}
where the symbol $\mu$ 
denotes the combination $(1+\alpha^2)\,\gamma^{-1}$.

Since the rhs of Eq.(\ref{llgs}) is orthogonal to the vector $\vec{m}$, 
its length is conserved, whereas  
direction varies in time. 
A characterization in terms of the spherical angles 
$\theta$ and $\varphi$ is suitable: $m_x=\sin\theta\cos\varphi$, 
$m_y=\sin\theta\sin\varphi$, $m_z=\cos\theta$, and the equations turn into
\begin{eqnarray}
\label{eq_angular}
 \mu\frac{d\theta}{dt}&=&(\alpha H_a-\overline{\beta} J M_0) 
         \cos\theta\cos\varphi\,
	 -( H_a+\alpha\overline{\beta} J M_0)\sin\varphi+\alpha S -T\\
 \mu\sin\theta\,\frac{d\varphi}{dt}&=&
 -(\alpha H_a-\overline{\beta} J M_0)\sin\varphi\,
 -(H_a+\alpha\overline{\beta} J M_0)\,\cos\varphi\cos\theta-S-\alpha T\nonumber,
\end{eqnarray}
where the symbols $S$ and $T$ denote, respectively, the combinations
$(H_{dz}+H_k\cos^2\varphi)\sin\theta\cos\theta$ and 
$H_k\sin\theta\sin\varphi\cos\varphi$.

Through the current $J$ these equations are coupled 
with dynamics of the LC-circuit. Electrodynamics of the latter
obeys the Kirchhof equation
\begin{equation}
%LC\,\ddot{V} +V =RJ=R(I-C\dot{V})
LC\,\frac{d^2V}{dt^2} +V =RJ=R\left(I-C\frac{dV}{Dt}\right)
\label{eq_circuit}
\end{equation}
where $V$ is the potential, $I$ is the ohmic current,
$L$ and $C$ are, respectively, inductance and capacitance of the circuit, 
and resistance of the circuit $R$ is the sum of the ohmic resistance
of the load and the resistance of the STO. The former is constant, but
the latter, due to the effect of magnetoresistance, is time-dependent: 
as shown in~\cite{Grollier_2006}, 
its value is a harmonic function of the instantaneous angle 
$\phi_{\vec{m}\vec{M}_0}$ between the magnetizations 
$\vec{m}$ and $\vec{M_0}$. 
The maximal value of resistance, $R_p$, is achieved in the case
when both magnetizations are parallel; the minimal one, $R_{ap}$,
corresponds to antiparallel magnetizations. 
In our configuration, $\vec{M}_0$ is directed along the $x$-axis,
hence $\phi_{\vec{m}\vec{M}_0}=\cos^{-1}(\sin\theta\cos\varphi)$.
Accordingly, 
\begin{equation}
R(\phi_{\vec{m}\vec{M}_0})=
\frac{R_p+R_{ap}}{2}\, (1-\varepsilon\sin\theta\cos\varphi)
\label{resistance}
\end{equation}
where $\varepsilon$ denotes the ratio $(R_p-R_{ap})/(R_p+R_{ap})$.

On combining (\ref{eq_angular}) with (\ref{eq_circuit}),
denoting the product $\overline{\beta} M_0$ by $\beta$
and absorbing $\mu$ in the time units, we arrive at the system 
of four ODEs~\cite{Pikovsky_2013};
\begin{eqnarray}
\frac{d\theta}{dt}&=&\cos\theta\cos\varphi\left(\alpha H_a-\beta I(1-w)\right)  
	 -\sin\varphi\left( H_a+\alpha\beta I(1-w)\right)
	   +\alpha S -T\nonumber\\
\label{eq_sto}
\sin\theta\,\frac{d\varphi}{dt}&=&
         -\sin\varphi\left(\alpha H_a-\beta I(1-w)\right)
	 -\cos\varphi\cos\theta\left(H_a+\alpha\beta I(1-w)\right)
	 -S-\alpha T\\
\frac{du}{dt}&=&\omega\, w\nonumber\\
\frac{dw}{dt}&=&\frac{\Omega_0^2}{\omega}\, 
       \bigg((1-\varepsilon\sin\theta\cos\varphi)\,(1-w)-u\bigg)\nonumber
\end{eqnarray}
where the variables $u$ and $w$ are, respectively, rescaled $V $ 
and $\dot{V}$, 
the abbreviations $S$ and $T$ denote, as above, the
combinations $(H_{dz}+H_k\cos^2\varphi)\sin\theta\cos\theta$ and 
$H_k\sin\theta\sin\varphi\cos\varphi$,
and the additional parameters which characterize the circuit are
$$ \omega=\frac{2(1+\alpha^2)}{\gamma\,(R_{p}+R_{ap})\,C}\;\;
\mbox{and}\;\;  \Omega_0^2=\frac{(1+\alpha^2)^2}{\gamma^2L\,C}.$$

These equations will be an object of our numerical investigations in the
following sections.
Recall that $I$ is the external current through the STO.
Below, in Sect.~\ref{sect_unforced} we treat $I$ as a constant;
further, in Sect.~\ref{sect_forced}, $I$ will be modulated around the
constant value with frequency ${\Omega}$ and amplitude $\eta$: 
$I(t)=I_0+\eta \cos{\Omega} t$. In this way, the equations for
both angular variables $\theta$ and $\varphi$ get affected by modulation.

Symmetry of the underlying physical problem finds reflection in the properties
of the dynamical system. External magnetic field $\vec{H_a}$ singles out in the physical
space a particular direction as well as the plane perpendicular to it. Accordingly, the
Cartesian equations (\ref{cartesian}) are invariant with respect to the simultaneous
change of sign of both components perpendicular to the field: $m_y$ and $m_z$.  
In terms of Eq.(\ref{eq_sto})
this means invariance with respect to the transformation 
$(\theta\to\pi-\theta,\;\varphi\to -\varphi)$. This symmetry, in turn, 
influences the properties of the solutions.

\section{Dynamics of a single autonomous STO}
\label{sect_unforced}
Consider an isolated spin-torque oscillator, governed by Eq.(\ref{eq_sto}).
Its phase space is a product of the sphere (angles $\theta$ and $\varphi$)
and the plane (variables $u$ and $w$). 
Below, we briefly delineate basic changes in dynamics, 
invoked by the increase of the current $I$.
We fix the values of other parameters at values taken in~\cite{Pikovsky_2013}: 
\begin{equation}
\alpha=0.01,\; \beta\,M_0=\frac{10}{3},\; \omega=1,\;\varepsilon=0.3,\;
H_{dz}=1.6,\, H_k=0.05,\; H_a=0.2. 
\label{parameters} 
\end{equation}
Concerning the remaining coefficient $\Omega_0$,  we
start with the fixed value $\Omega_0$=1/2 used in~\cite{Pikovsky_2013}, 
(there this coefficient was called $\Omega$),
but later switch to $\Omega_0$=3/2; 
the reason will be explained below.

\subsection{Equilibria and their stability}
A magnetic moment $\vec{m}$,  set parallel to the external field $\vec{H}_a$,
preserves its direction forever and stays constant; 
the off-field components will not be excited, and precession would not arise.
There are two possibilities: $\vec{m}$ is  directed either along $\vec{H}_a$ 
or in the opposite direction.
In the chosen Cartesian
coordinates, such states are characterized  by $m_x=\pm 1$
whereas $m_y$ and $m_z$ vanish identically.
Setting in Eq.(\ref{eq_sto}) $\sin\varphi$=0, $\cos\theta=\pm$1 yields these
two equilibria which exist at all parameter values.
Difference in the sign of the product $\cos\varphi\,\sin\theta$ 
between these states finds reflection in 
the difference of their stability properties.
\subsubsection{Quiescent state}
We start with the ``natural'' state of equilibrium in which the magnetic moment
is aligned with the external field.
Coordinates of this state in the phase space are: 
\begin{equation}
\theta=\frac{\pi}{2},\;\varphi=0,\;
%\mbox{(or, equivalently,}\; \theta=-\frac{\pi}{2}, \varphi=\pi\,),\;
u=1-\varepsilon,\; w=0.
\label{quiescent}
\end{equation}
Characteristic equation of its Jacobian is the product 
of two second order polynomials.
The first one, independent of $\alpha,\beta,H_a,H_k,H_{dz}$ and the current $I$,
yields two eigenvalues 

\begin{equation}
\lambda=\frac{\Omega_0}{2\omega}\,\left(\Omega_0(\varepsilon-1)
   \pm\sqrt{\Omega_0^2(\varepsilon-1)^2-4\omega^2}\right) 
\label{qui_lambda_12}
\end{equation}
For $\Omega_0<2\omega/|\varepsilon-1|$ the latter are complex, and, since
$\varepsilon<1$, possess negative real parts.
The second factor is 
\begin{equation}
\lambda^2+2\lambda\left(\alpha (H_a+H_k+\frac{H_{dz}}{2})-\beta I\right)
+(1+\alpha^2)\left(\beta^2 I^2+(H_a+H_k)(H_a+H_k+H_{dz})\right).
\label{qui_lambda_34}
\end{equation}

Since the last term is positive, these two eigenvalues cannot have 
opposite signs.
If the current $I$ is absent or sufficiently weak, the equilibrium is stable; 
it gets destabilized in the Hopf bifurcation at
\begin{equation}
I=I_H=\frac{\alpha}{\beta}(H_a+H_k+\frac{H_{dz}}{2}).
\label{bif_Hopf}
\end{equation}
For the employed parameter values, this $\omega$-,$\Omega_0$- and 
$\varepsilon$-independent threshold lies at $I_H=0.00315$.

\subsubsection {Saddle equilibrium}
Another equibrium state corresponds to the situation where
the magnetic moment is directed opposite to the external field.
It is natural to expect that such a state would be unstable;
however, this is not necessarily the case.
For this steady state  with coordinate values
\begin{equation}
\theta=\frac{\pi}{2},\;\varphi=\pi,\;
%\mbox{(or, equivalently,}\; \theta=-\frac{\pi}{2},\,\varphi=0),\; 
u=1+\varepsilon,\; w=0,
\label{saddle}
\end{equation}
the characteristic equation consists of two factors as well.
The first one is the polynomial 
\begin{equation}
\lambda^2+\lambda\frac{\Omega_0^2}{\omega}(1+\varepsilon)+\Omega_0^2
\label{char_saddle_1}
\end{equation}
whose roots always possess negative real parts. 
For $\Omega_0>2\omega/(\varepsilon+1)$ 
the roots are real; otherwise they are complex. 
Notably, the corresponding eigenvectors in the phase space
are orthogonal to the plane of angular variables $\theta$ and $\varphi$.

The second factor of the characteristic equation is 
\begin{equation}
\lambda^2+2\lambda\left(\beta I+\alpha (H_k-H_a+\frac{H_{dz}}{2})\right)
+(1+\alpha^2)\left(\beta^2 I^2+(H_k-H_a)(H_k-H_a+H_{dz})\right).
\label{char_saddle_2}
\end{equation}
For $\displaystyle I<I_{st}=\frac{1}{\beta}\sqrt{(H_a-H_k)(H_k-H_a+H_{dz})}$, 
the last term is negative, and the steady state is a saddle. 
At $I=I_{st}$ it gets stabilized in the course of the pitchfork bifurcation. 

For the employed values of $\beta,H_a,H_k$ and $H_{dz}$, \ $I_{st}=0.1399$, 
and by far exceeds $I_H$. 
However, through variation of parameters (e.g. increase of $\alpha$, or, 
conversely,  choosing close values of $H_a$ and $H_k$) 
the regions of stability of the two steady states 
can be made closer and even brought to an overlap. 

In the case when the equilibrium (\ref{saddle}) is unstable, 
the eigenvector which corresponds to the positive eigenvalue 
lies in the plane of coordinates $\theta$ and $\varphi$.
Hence, the coordinate transformation 
$(\theta\to\pi-\theta,\;\varphi\to -\varphi)$ interchanges 
two components of the unstable manifold. 

\subsubsection {Further steady states}
Besides (\ref{quiescent}) and (\ref{saddle}), equations (\ref{eq_sto}) 
possess other steady solutions. Two of them, in absence of external current $I$, 
are characterized by 
$m_x=-H_a/(H_{dz}+H_k),\,m_y=0,\,m_z=\pm\sqrt{1-m_x^2},\;u=1-\varepsilon m_x$; 
explicit expressions for their coordinates at $I\neq 0$ are too long to be quoted here.
%$\sin\theta=-H_a/(H_{dz}+H_k),\,\varphi=0$
At the employed values of the parameters the corresponding fixed points are 
unstable foci located  in the polar regions of the sphere spanned by $\theta$ and 
$\varphi$. They do not play immediate role in the dynamics, 
but are helpful in classifying the periodic solutions of Eq.(\ref{eq_sto}). 

Outside the considered parameter range,
further steady solutions may exist,  
e.g. symmetric branches born from (\ref{saddle}) at $I=I_{st}$; they are not relevant
for the dynamics that we describe below.

\subsection{Peculiarities of homoclinic bifurcations}

The supercritical Hopf bifurcation at $I_H$ gives birth to the small-scale
periodic oscillations of the magnetic moment. 
When the current $I$ is further increased, the amplitude of oscillations, 
as well as their temporal period grow, 
the corresponding phase portrait approaches the saddle point, 
and at a certain critical value  $I=I_{\rm hom}$ it gets captured by 
the unstable manifold of the saddle: the homoclinic bifurcation takes place. 
This event marks the transition from oscillations with bounded ($<2\pi$) 
amplitude with respect to $\varphi$ to unbounded rotations along this 
angular variable. Although this transition produces neither a topological 
change in the type of the closed phase trajectory on the sphere spanned 
by $\varphi$ and $\theta$, nor a  noticeable change in the characteristics 
of the relevant physical observable 
($x$-component of the magnetic moment $m_x$), 
below for clarity we distinguish the ``pre-homoclinic'' $\varphi$-oscillations
 from the ``post-homoclinic'' $\varphi$-rotations.

\begin{figure}[h]
\centerline{\includegraphics[width=0.6\textwidth]{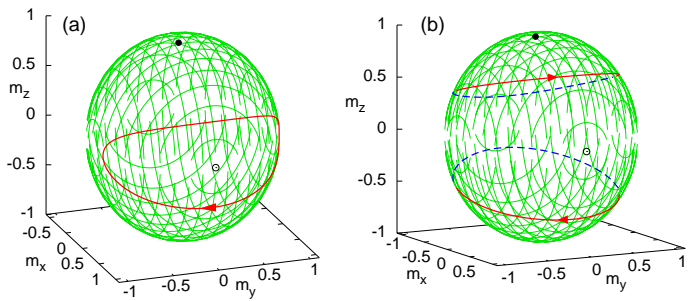}}
\caption{Closed orbits of Eq.(\ref{eq_sto}) on the spherical surface.
(a) $\varphi$-oscillations ($I$=0.0035); (b): $\varphi$-rotations ($I$=0.008).\protect\\
Empty circle: equilibrium  (\ref{quiescent}).  Filled circle:
unstable steady state in the polar regions.
}
\label{spheres}
\end{figure}

Distinction between these two kinds of periodic solutions can be conveniently
expressed in terms of location of the corresponding phase curves 
with respect to the fixed points on the surface of the sphere.
Altogether, at $I>I_H$ there are four steady states: 
three unstable focus points and a saddle.  
A closed phase curve divides the surface of the sphere into two parts:
the part with a single equilibrium of the focus type and the part
with three remaining equilibria. We say that the single
equilibrium is ``encircled'' by the curve. It is straightforward
to see (cf Fig. \ref{spheres}) that the closed curve 
of  $\varphi$-oscillations encircles the steady state (\ref{quiescent}) 
whereas the closed curve of  $\varphi$-rotations encircles one
of the near-polar foci. In this sense, oscillations characterize precession
around the axis of the applied magnetic field whereas rotations 
correspond to precession with the axis close to the direction of demagnetization.

Symmetry of equations transforms each component of the unstable
manifold of the saddle point into the other component. Therefore, homoclinic
orbits occur in symmetric pairs: as a result, homoclinic bifurcations
are accompanied in the phase space not by birth/destruction of closed orbits, 
as in the conventional homoclinic setup, but by their recombination 
(``gluing''), so that in the parameter space closed trajectories exist 
both before and after homoclinicity. The consequences of formation and breakup 
of homoclinic  orbits depend on the type and relative magnitudes 
of the eigenvalues of the saddle. As proven in the seminal papers of 
Shilnikov~\cite{Shilnikov_63,Shilnikov_68}, in the situation
when the Jacobian at the saddle point has a single positive eigenvalue
$\lambda_+$, the decisive part is played by the 
sum $\lambda_++{\rm Re}(\lambda_-)$
where $\lambda_-$ is the eigenvalue with
the closest to zero negative real part. If this sum
(often called the ``saddle quantity'') is negative
-- in other words, if in the vicinity of the
saddle contraction prevails over expansion, -- 
the breakup of the homoclinic trajectory 
leaves in the phase space the unique stable periodic orbit. 
If the saddle quantity is positive, the periodic orbit is unstable. 
Moreover, if in the latter case  $\lambda_-$
is complex, the system possesses a countable set of unstable periodic orbits
which, in many situations, serves as an indicator of chaotic dynamics.

\begin{figure}[h]
\centerline{\includegraphics[width=0.8\textwidth]{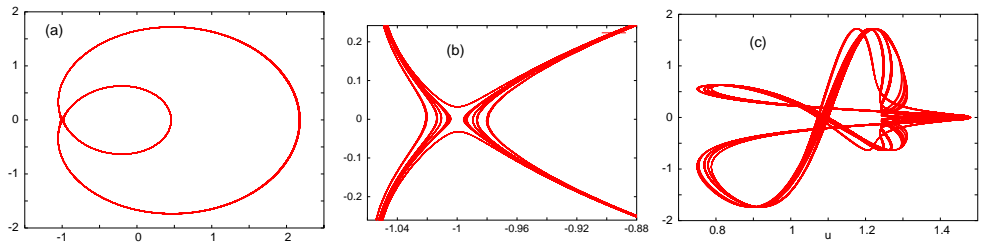}}
\caption{Projections of attractor at $\Omega_0$=0.5, $I$=0.00568,
other parameters see (\ref{parameters}). 
(a),(b): stereographic coordinates,  
$\cos\varphi\tan\left(({\pi-2\theta})/{4}\right)$ vs.
$\sin\varphi\tan\left(({\pi-2\theta})/{4}\right)$.
(c): $u$ vs. $\sin\varphi\tan\left(({\pi-2\theta})/{4}\right)$. }
\label{fig_2}
\end{figure}

Let us start with the case of $\Omega_0=1/2$.
The $I$-independent complex eigenvalues from Eq. (\ref{char_saddle_1}) 
equal $-0.1625\pm0.4729\,\rm i$. 
In the parameter range $0.005<I<0.006$ where the homoclinic
bifurcation is encountered, the real eigenvalues from (\ref{char_saddle_2}) 
are $\approx$0.44 and $\approx$~--0.49.  Accordingly, the steady state is a
saddle-focus: the leading direction on its stable manifold corresponds to 
complex eigenvalues. Remarkably, since this direction is orthogonal 
to the plane of $\phi$ and $\theta$, 
the oscillatory stage in the approach to the steady state is almost invisible  
from the point of view of angular variables, as well as
upon a stereographic projection.  The fine structure, hardly visible in
stereographic coordinates (left panel of Fig.~\ref{fig_2}) can be resolved
with a blowup (middle panel),  but is well recognizable in terms of the variable
$u$ (right panel). 

Since the repulsion along the unstable manifold is faster 
than the contraction along the leading direction of the stable one, 
the saddle quantity is positive, and existence of the 
pair of homoclinic trajectories 
implies presence of countably many unstable periodic orbits, horseshoes etc. 
Moreover, since the sum of three leading eigenvalues is positive, there should 
be no stable periodic solutions in the vicinity of homoclinic 
orbits~\cite{Shilnikov_book}.

Apparently, the stable periodic solution with bounded $\varphi$, 
born in the Hopf bifurcation and numerically observed until $I$=0.005688, 
does not directly participate in the homoclinic bifurcation.
It comes reasonably close to the saddle point, so that its period grows
from 27.045 at $I$=0.005 to 39.885 at $I$=0.005688,
but disappears in the saddle-node bifurcation.

\begin{figure}[h]
\centerline{\includegraphics[width=0.8\textwidth]{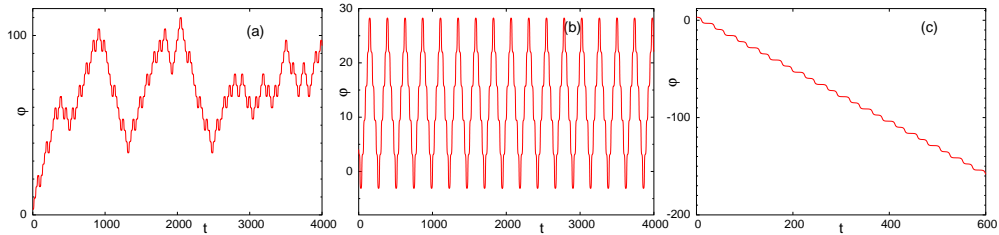}}
\caption{Temporal evolution of $\varphi$ close to homoclinics.
$\Omega_0$=0.5, other parameters as in (\ref{parameters}).
(a): $I$=0.00567; (b) $I$=0.00568; (c)  $I$=0.0057.}
\label{fig_3}
\end{figure}

Tracking the unstable manifold of the saddle point indicates that homoclinics 
occurs slightly before that event, at $I$=0.005624$\ldots$
The (first) homoclinic bifurcation 
and the disappearance of the stable limit cycle with bounded oscillations 
of $\varphi$ do not yet signify the transition to monotonic 
$\varphi$-rotations: immediately after it, (e.g. at $I$=0.00567) 
clockwise and counterclockwise rotations 
alternate in a seemingly irregular way, as shown in the left panel 
of Fig.~\ref{fig_3}.
Further from the bifurcation point (e.g. at $I$=0.00568) the trajectory 
makes a few consecutive rotations clockwise, 
followed by the same number of counterclockwise rotations,
so that the angle $\varphi$ remains confined. 
Finally, starting from $I\approx 0.0057$, the monotonic drift of $\varphi$ 
begins.  

Altogether, complicated dynamics is restricted to a small parameter range: 
$0.0055<I<0.0058$; outside it, attractors are periodic and 
have simple structure.
Below, we do not elaborate on the aspects of chaotic dynamics. 
On the contrary, we would like to avoid complications associated 
to presence of countably many periodic orbits in the phase space
and to deal with the evolution of just one periodic solution.
For this purpose, at the event of homoclinic bifurcation 
both leading eigenvalues of the saddle state should be real. 
This can be achieved e.g. by varying the parameter $\Omega_0$
which influences the complex eigenvalues but leaves the real ones intact.  
When $\Omega_0$ is increased beyond $\approx$0.62, contraction along the stable 
manifold becomes {\em less than twice as fast} as contraction along the stable 
one, hence close to homoclinicity stable periodic orbits should be dense in the 
parameter space~\cite{Shilnikov_book}. 
Above $\Omega_0 \approx$ 0.83 the absolute value 
of the smallest negative real part of the eigenvalue 
exceeds the positive eigenvalue: the saddle quantity turns negative, 
and the homoclinic bifurcation gives rise to stable periodic solutions.
Further on, with $\Omega_0$ above $\approx$0.85,
the smallest negative real part belongs not to the complex pair but to
the real negative eigenvalue which corresponds to angular variables. Hence, 
the saddle-focus turns into the saddle point, with all usual implications for
homoclinic bifurcations. In contrast to the case of the saddle-focus 
with its intermediate large scale oscillations, 
the homoclinic bifurcation with the saddle configuration
immediately leads to monotonic evolution of $\varphi$.
Since the leading direction on the stable manifold
belongs to the plane of angular variables, 
two components of the unstable manifold approach the saddle 
from the opposite directions (so-called \textit{number eight} configuration, 
opposed to the \textit{butterfly}, familiar from the pictures of the homoclinic
bifurcation in the Lorenz equations). 

\begin{figure}[h]
\centerline{\includegraphics[width=0.8\textwidth]{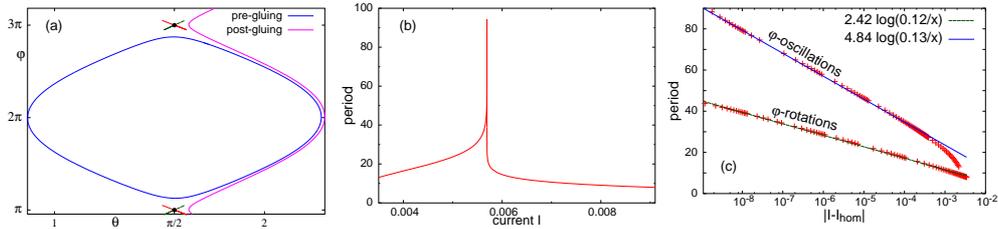}}
\caption{Gluing bifurcation. 
(a): transformation of the attracting orbit,
(b),(c): parameter dependence of period of oscillations
close to the bifurcation point.}
\label{fig_4}
\end{figure}

It is convenient to view the transformation of the trajectory in terms
of the variable $\varphi$ unwrapped onto the line: 
the homoclinic orbit turns into the saddle connection 
between two $2\pi$-shifted replicas of the saddle point
(left panel of  Fig. \ref{fig_4}).
Take a segment of the attracting trajectory which departs from the saddle 
along, say, its right separatrix, and therefore approaches the next 
(shifted) saddle from the right.
Before the bifurcation, it turns back and returns to its initial position 
along the left separatrix, producing  a closed trajectory.
In contrast, after the bifurcation the same trajectory 
always passes the saddles 
``from the right'', ensuring unbounded growth of $\varphi$: 
the clockwise $\varphi$-rotation. 
Symmetric counterpart of this orbit describes the counterclockwise rotation.
Up to non-conservation of energy, qualitatively this is the same 
heteroclinics-mediated transition 
from oscillations to rotations as in the classical nonlinear pendulum 
$\ddot{\varphi}+\sin{\varphi}=0$.

Finally, for $\Omega_0>2\omega/(\varepsilon+1)$=1.538, 
all eigenvalues of the saddle steady state become real, without visible
consequences for dynamics.

Variation of $\Omega_0$ does not affect the threshold $I_H$ 
of the onset of oscillations.
A bit surprisingly, it also almost does not affect the value of $I$ 
which corresponds to the homoclinic bifurcation: 
for the configuration with $\Omega_0=0.5$ 
the latter occurs at $I=0.0056244\ldots$
whereas for $\Omega_0=1.6$ it takes place at $I=0.0057098\ldots$. 

To ensure the uniqueness of the periodic orbit which participates in
the homoclinic bifurcation, below we use the
parameter value $\Omega_0=1.5$. When the values of the other parameters 
are fixed according to  (\ref{parameters}), 
the homoclinic bifurcation occurs at $I_{\rm hom}$=0.0056929063.
Close to the bifurcation, the period of the orbit logarithmically 
diverges (central panel of Fig. \ref{fig_4}):
$T\sim -A_{\rm o,r} \log|I-I_{\rm hom}|$.
As seen on the left panel, during each oscillatory cycle 
the trajectory twice hovers in the neighborhood of the saddle point, 
whereas a cycle of $\varphi$-rotations includes
only one passage close to the saddle.
Therefore, the prefactor $A_o$ on the side of oscillations is twice 
as large as the prefactor $A_r$ on the side of rotations (right panel).
Non-monotonic dependence of the period on $I$ leads to the remarkable
consequence for the case of periodic driving: for a given value 
of the driving frequency, there are two ranges of $I$ where
phase locking should be encountered.

We expect that strong variation of period within a small parameter interval 
should affect the sensitivity of oscillations to the external periodic signal.

%%%%%%%%%%%%%%%%%%%%%%%%%%%%%%%%

\section{Return mapping and basic bifurcations in the case
of external modulation} 
\label{sect_mapping}
Now we introduce periodic modulation of the current in Eq.(\ref{eq_sto}):
$I(t)=I_0+\eta \cos{\Omega} t$. For simulations we take
the frequency ${\Omega}$ which is of the order of the frequency 
of self-sustained oscillations in the autonomous system, as well as
the values of $I_0$ above the threshold $I_H$, and the values
of amplitude $\eta$ which do not exceed $I_0$.
Numerical integration shows that phase portraits 
of oscillatory states include segments of 
trajectories which pass relatively close to the location of the 
saddle equilibrium in the autonomous system.
 
Let us construct and analyze a simplified mapping which
interrelates the coordinates of subsequent returns into this region.
We restrict ourselves to the case when in the spectrum of the saddle,
the eigenvalue $\lambda_-$ with the smallest negative real part is real;
in Eq.(\ref{eq_sto}) this corresponds to $\Omega_0>0.85$.

\subsection{Autonomous flow}
\begin{figure}[h]
\centerline{\includegraphics[width=0.8\textwidth]{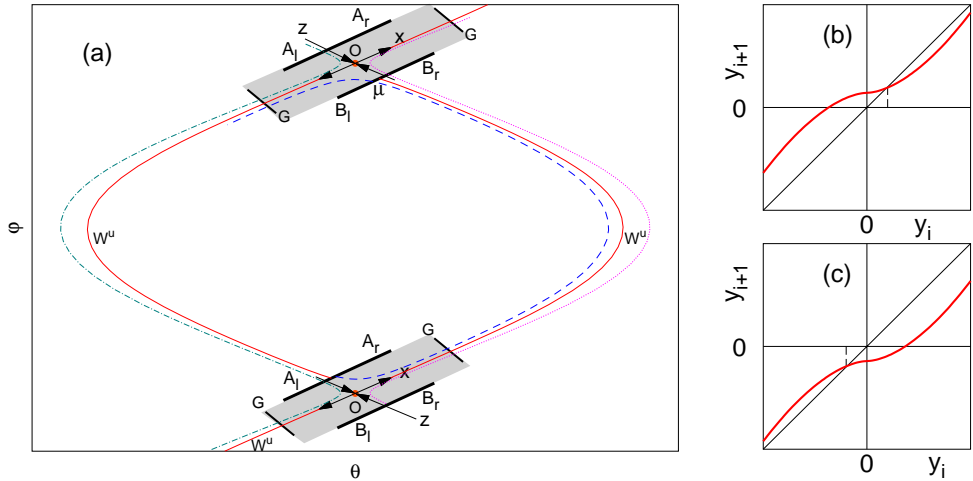}}
\caption{Return mapping near the saddle point.
(a): Construction of the mapping. 
Gray domains: regions of local (linearized) dynamics. 
$O$: copies of the saddle point; red curves $W^u$: components
of the unstable manifold; 
$A_{l,r},B_{l,r}$: Poincar\'e secants, transversal to the stable
manifold; $G$: secants, transversal to $W^u$; 
dashed and dotted lines: typical phase trajectories
passing through the vicinities of the saddle.\protect\\
(b),(c): One-dimensional map (\ref{combined_map}) with $\nu>1$ 
at, respectively, negative and positive $\mu$.}
\label{fig_5}
\end{figure}

To start with, we review the return mapping in the autonomous situation.
We do not discriminate between an autonomous saddle point 
(say,  $\theta=-\pi/2,\,\varphi=0$), 
and all its replicas shifted by $2\pi n\; (n=\pm1,\pm2,\ldots)$
along $\varphi$. 
However, in order to capture the difference between $\varphi$-oscillations 
and $\varphi$-rotations,
we need to consider the region of phase space with two such replicas 
(see Fig.\ref{fig_5}).

We parameterize directions along the unstable and stable
manifolds of the saddle point by local coordinates $x$ and $z$, 
respectively\footnote{this notation is not related to orientation
of the magnetization vector from Eq.(\ref{llgs}).}.
Two components of the unstable manifold $W^u$ (separatrices) 
are transformed into each other by
the simultaneous change of sign of $x$ and $z$. The same transformation
maps every orbit of the system either onto itself, 
as in the case of periodic oscillations, or onto its symmetric image 
(e.g. clockwise and counterclockwise rotations).

Near the saddle point, the equations of motion are 
\begin{equation}
dx/dt=\lambda_+ x + h.o.t.,\;\;
dz/dt=\lambda_- z + h.o.t.
\end{equation}
where $\lambda_+$ and $\lambda_-$ are, respectively, the positive
and the closest to zero negative eigenvalues of the Jacobian at the saddle.
The usual formalism divides the phase space into the ``local'' part 
near the saddle (shown as gray polygon in Fig.\ref{fig_5}, 
and the ``global' region far from the fixed point(s)
where the trajectories stay close to the separatrices. 
Trajectories enter the local region by crossing either of two lines $A,B$, 
transversal to the stable manifold of the saddle, and exit by crossing
either of  two lines $G$, transversal to its unstable manifold. 
These borders are taken sufficiently close to the saddle, 
so that the flow in the local region can be viewed as linear.
The Poincar\'e sections are taken at the entrance to the local
region, on both sides of the saddle,
at small distances $\pm z_s$ from its local unstable manifold. 
An orbit which starts from either of the secants $A,B$
with a positive (negative) value of $x$,
departs from the saddle along the ``right'' (``left'') 
separatrix, and arrives to the region of the saddle replica,
shifted with respect to $\varphi$ by $2\pi$ ($-2\pi$). 
Accordingly, we divide each secant into the left and right parts
$A_l$ and $A_r$, respectively $B_l$ and $B_r$.
A section is characterized by the coordinates $x_i$ and $z_i$
of the intersection point; in fact, for the latter only the sign matters.
In the global region, trajectories stay close to the separatrices 
of the saddle. 

This approach allows to 
interrelate the coordinates of consecutive arrivals on Poincar\'e secants; 
in the lowest relevant order these relations read
\begin{equation}
x_{i+1}=\mu+a x_i^\nu {\rm sign}(z_i),\;\; z_{i+1}=-z_s
\label{right_map}
\end{equation}
for orbits with positive $x_i$ 
which move along the right separatrix, and 
\begin{equation} 
x_{i+1}=-\mu-a |x_i|^\nu {\rm sign}(z_i),\;\; z_{i+1}=z_s 
\label{left_map}
\end{equation}
for orbits with negative $x_i$ which depart along the left separatrix.
Here, $\mu$ is the $x$-coordinate of the first intersection 
of the right separatrix with the Poincar\'e secant $z=-z_s$; 
the value $\mu$=0 corresponds to formation of the homoclinic trajectory. 
Variation of the parameter $\mu$  roughly reflects 
variation of the current $I_0$ in the original equations.
Further, the saddle index $\nu=|\lambda_-|/\lambda_+$ 
shows which process -- contraction or expansion --
prevails in the vicinity of the saddle point. 
According to the Shilnikov theory,
values of $\nu$ between 0 and 1 ensure instability of periodic orbits, born
from the breakup of the homoclinic trajectory. 
For $\nu>1$ such orbits are, on the contrary, stable; this corresponds to 
the relevant region of the parameter space of Eq.(\ref{eq_sto}), where
the value of $\nu$ lies between 1.6 and 1.7. 
Finally, the factor $a$ is the so-called ``separatrix value''; its sign shows 
whether the mapping along the separatrix preserves orientation. 
Numerical simulation of Eq.((\ref{eq_sto}) indicates 
that in our situation (which is reminiscent of dynamics 
on the two-dimensional sphere) the mapping is orientable,
hence $a$ is positive; below we set $a=1$ (at $\nu\neq 1$ this is achieved
by rescaling the units of $x$). 

To ensure validity of the mapping, the system should stay sufficiently close
to homoclinicity, and the considered trajectories should be selected 
not too far from the separatrices. 
Expressed in terms of $\mu$ and $y$, this reads 
$|\mu|\ll (\nu-1)\,\nu^{\nu/(1-\nu)}$
 and $|y|< \nu^{1/(1-\nu)}$.

To unify two mappings (\ref{right_map}) and (\ref{left_map}), we introduce
the auxiliary variable $y=x\,{\rm sign}(z)$\cite{Lyubimov_Byelousova}; $y$ 
is invariant with respect to the simultaneous change of signs of $x$ and $z$.
In terms of $y$, the return mapping reduces to 
\begin{equation}
y_{i+1}=-\mu+ |y_i|^\nu\, {\rm sign}(y_i)
\label{combined_map}
\end{equation}

Plots of the  piecewise-smooth mapping (\ref{combined_map}) for $\mu<0$
and $\mu>0$ are shown in Fig.\ref{fig_5}(b,c). 
At small negative values of $\mu$ the mapping
possesses a stable fixed point which lies close to zero at $y>0$; at small
positive values of $\mu$, the value of $y$ in the fixed point 
is negative. While interpreting this in terms of the original
variables, we should keep in mind the assignment of signs in $y$.
The fixed point with positive $y$ corresponds e.g.
to passage from $A_r$ to $B_l$ in Fig.~\ref{fig_5}; 
since there the hitherto ascending trajectory turns down and returns to the
lower saddle, the iteration of the mapping corresponds to the half of the
oscillation of $\varphi$; the complete cycle of 
oscillation requires two iterations. In contrast, 
the fixed point with negative $y$ corresponds to the passage of the
trajectory from $A_r$ upwards to $A_r$ 
or from $B_l$ downwards  to $B_l$. 
With each new iteration this passage is repeated, hence the fixed point with
negative $y$ describes the clockwise and the counterclockwise rotations of
the variable $\varphi$. Since the mapping (\ref{combined_map}) is monotonic,
there are no other local attractors. For completeness we also mention two
distant unstable fixed points of the mapping, with large values of $|y|$ 
(at $\mu=0$ they lie at $y=\pm 1$); the point with positive $y$ may
be vaguely associated with the equilibrium (\ref{quiescent}), destabilized 
in the course of the Hopf bifurcation.
 
\subsection{Return mapping under periodic forcing}

With this picture in mind, let us introduce 
a sufficiently weak periodic force acting on both
leading variables. The forced equations of motion read
\begin{eqnarray}
dx/dt&=& \lambda_+ x+ A\cos\omega t+h.o.t.\nonumber\\
dz/dt&=& \lambda_- x+A\cos\omega t +h.o.t. \nonumber
\label{2_a}
\end{eqnarray}
When the nonlinear terms are discarded, the solutions
are $x(t)=c_1{\rm e}^{\lambda_+ t} + x_p(t)$ and 
$z(t)=c_2{\rm e}^{\lambda_- t} + z_p(t)$ where $x_p(t)$ and $z_p(t)$ are
components of the unstable periodic solution
$$x_p(t)=
   \frac{A}{\omega^2+\lambda_+^2}(\omega\cos\omega t-\lambda_+\sin\omega t) ,\; 
   z_p(t)=
   \frac{A}{\omega^2+\lambda_-^2}(\omega\cos\omega t-\lambda_-\sin\omega t) .$$

On rewriting the equations in the corotating reference frame by
introducing the variables $\tilde{x}\equiv x-x_p(t)$ and 
$\tilde{z}\equiv z-z_p(t)$, they become
\begin{equation}
d\tilde{x}/dt=\lambda_+ \tilde{x} + F_1(\tilde{x},\tilde{z},t),\;\;
d\tilde{z}/dt=\lambda_- \tilde{z} + F_2(\tilde{x},\tilde{z},t)
\label{corotating}
\end{equation}
where $F_1$ and $F_2$ consist of nonlinear terms with explicit periodic
dependence on $t$. In this sense, additive periodic forcing turns into
periodic modulation of the parameter(s).

Equations (\ref{corotating}) 
have an equilibrium point at the origin: $\tilde{x}=\tilde{z}=0$. 
For the derivation of the return mapping, the above formalism
should be extended in order to account for the
variability of the force at the moments of intersection with
the Poincar\'e secants. As a coordinate characterizing the external forcing, 
it is convenient to take the phase $\Psi$ of the force; it
obeys the equation $d\Psi/dt=\omega$. For the given orbit, 
let $\tau_{\rm loc}$ and $\tau_{\rm g}$ be the times of passage 
of the trajectory through, respectively, the local and the global region.
Then, the increment of the phase $\Psi$ between
subsequent intersections equals
$\omega(\tau_{\rm loc}+\tau_{\rm g})$. 
The value of $\tau_{\rm g}$, in general, depends on the values 
of $\tilde{z}$ and $\Psi$ at the exit from the local region;
to start with, we neglect this
dependence and treat $\tau_{\rm g}$ as a constant. 
The local passage time $\tau_{\rm loc}$ depends on the place of entry to the
local region; up to an additive constant (which 
can be absorbed into $\tau_{\rm g}$
or, by rescaling $x$, can be turned into zero),
it equals $-\lambda_+^{-1}\log|\tilde{x}_i|$. 

We expect that effect of the weak forcing upon the section
coordinate $\tilde{x}_i$
can be represented as an additive modulation, proportional to the
value of the force at the exit from the local region. 
In this way, we arrive at the mapping which relates the values 
of $y$ (recall that $y=\tilde{x}\,{\rm sign}(\tilde{z})$) 
and $\Psi$ at the moments
of subsequent intersections with the Poincar\'e planes:
\begin{eqnarray}
\label{map_2D}
y_{i+1}&=&-\mu+ |y_i|^\nu\, {\rm sign}(y_i)
       +B\sin(\Psi_i-\omega\lambda_+^{-1}\log|y_i|)\\
\Psi_{i+1}&=&\Psi_i
     +\omega(-\lambda_+^{-1}\log|y_i|+\tau_{\rm g})\nonumber
\end{eqnarray}
where $B$ is proportional to the forcing amplitude $A$.

The mapping of this kind and its dynamics were described in~\cite{Afr_Shil}; 
there, the frequency of the forcing was not explicitly included 
in the analysis.
In our setup, the driving frequency $\omega$ plays the crucial role, 
and we treat it as an independent parameter. 
A similar mapping was partially analyzed in \cite{ZPK} in the
context of chaotic synchronization; there, the case of $\nu<1$ was considered, 
so that the corresponding periodic orbit in the autonomous system was unstable.

A periodic solution of the continuous system  
corresponds to the fixed point of the mapping (\ref{map_2D}): 
$y_{i+1}=y_i$ and $\Psi_{i+1}=\Psi_i+2\pi$.
The latter relation yields two values for the coordinate $y$ 
of the fixed point:
\begin{equation}
y=\pm\exp\bigg(\lambda_+(\tau_{\rm g}- \frac{2\pi}{\omega})\bigg).
\label{y_fp}
\end{equation}
Recall that the fixed point with positive $y$ corresponds to the
$\varphi$-oscillations whereas the fixed point with $y<0$ describes
the rotational state.
Each value of $y$, in its turn, leads to two values of the phase of the
force $\Psi$:
\begin{equation}
\Psi=\omega\tau_{\rm g} +
\left\{\begin{array}{c} \displaystyle\arcsin\frac{\mu+y-|y|^\nu  
       {\rm sign}(y)}{B}\\
\displaystyle\pi-\arcsin\frac{\mu+y-|y|^\nu {\rm sign}(y)}{B}\end{array}\right. 
\label{psi_fp}
\end{equation}
For the analysis of existence and stability of the fixed point we take
more general functional dependencies:
\begin{eqnarray}
y_{i+1}&=&f(y_i)+B g(\Psi_i+\omega\tau_{\rm loc}(y_i))\\\
\Psi_{i+1}&=&\Psi_i
     +(\tau_{\rm loc}(y_i)+\tau_{\rm g})\,\omega\nonumber
\label{map_general}
\end{eqnarray}
using, along
with the 2$\pi$-periodic function $g$, 
the characteristics of the autonomous system: 
the return mapping $f(y)$  and the passage time $\tau_{\rm loc}(y)$.

The saddle-node bifurcation in this system takes place at
$$ g'(\phi)\tau'_{\rm loc}(y)=0;$$
here and below $\phi$ is the abbreviation for the argument of $g$,
and all derivatives are taken at the fixed point. 
Since the local passage time
$\tau_{\rm loc}(y)$ has no extrema, this condition reduces to $ g'(y)=0$.

The period-doubling bifurcation requires
$$ 2+2f'(y)+B\,g'(\phi)\omega\tau'_{\rm loc}(y)=0.$$
Finally, the torus (Neimark-Sacker) bifurcation occurs at
$$f'(y)=1.$$

Substituting the particular dependencies $f,g$ and $\tau_{\rm loc}(y)$,
we obtain explicit expressions for the stability boundaries.

\subsubsection{Oscillations}

On the parameter plane of $\omega$ and $B$, the pair of fixed points 
(\ref{psi_fp}) with positive $y$ exists 
above the curve of the saddle-node bifurcation
\begin{equation}
\label{sb_plus} 
B_{\rm sn}^{\rm o}(\omega)
     =|\,\mu+{\rm e}^{\lambda_+(\tau_{\rm g}-2\pi/\omega)}
          -{\rm e}^{\nu\lambda_+(\tau_{\rm g}-2\pi/\omega)}\,|
\end{equation}
which has the ``infinitely flat'' horizontal asymptote at $\omega\rightarrow 0$.

At $\mu<0$, when the oscillatory solution is present already 
in the autonomous system, the region of existence (Arnold tongue) 
opens up from the point at the abscissa $B=0$; the value of $\omega$ in
this point equals the frequency $\omega_{\rm aut}$ of autonomous oscillations.
On the left boundary of the tongue, 
$$B_{\rm sn}^{\rm o}(\omega)=-\mu-{\rm e}^{\lambda_+(\tau_{\rm g}-2\pi/\omega)}
                     +{\rm e}^{\nu\lambda_+(\tau_{\rm g}-2\pi/\omega)},$$
at very small $\omega$ the asymptote is approached from below; note
that two $\omega$-dependent terms contribute with different signs.	    
		     
In contrast, at $\mu>0$, when the oscillatory solution
is absent in the autonomous system, the expression (\ref{sb_plus}) 
is bounded away from zero at small and moderate values of $\omega$. 
Here, an oscillation 
can be excited by the periodic driving force, 
provided that the amplitude $B$ exceeds the frequency-dependent
threshold value and, in any case, is larger than $\mu$.

Just above the saddle-node bifurcation, one of the newborn
fixed points is stable (the second one is a saddle). 
The region of stability is bounded from above
by the curve of the period-doubling bifurcation $B_{\rm pd}(\omega)$. 
It is convenient to characterize this curve in terms of its elevation
with respect to $B_{\rm sn}(\omega)$:
\begin{equation}
\label{pd_osc} 
B_{\rm pd}^2 (\omega)- B_{\rm sn}^2(\omega) 
   =\frac{4\lambda_+^2 y^2}{\omega^2}\left(1+\nu|y|^{\nu-1}\right)^2
\end{equation}
Since $y$ is exponentially small with respect to $\omega$
(cf. Eq.(\ref{y_fp})), the stability region becomes very narrow and, 
practically, nearly vanishes at low values of the driving frequency. 
Remarkably, however,
two bounding curves do not intersect at finite $\omega$.

It should be recalled that the fixed point of the mapping (\ref{map_2D})
with positive $y$ corresponds not to the whole cycle of oscillations,
but to half of that cycle; in the flow, it describes symmetric phase curves 
which are mapped onto themselves by transformation $\varphi\to - \varphi$,
$\theta\to \pi/2-\theta$. Accordingly, the period-doubling bifurcation
in (\ref{map_2D}) will, in terms of the flow, be not the period-doubling,
but the symmetry-breaking (pitchfork) bifurcation of the limit cycle.

The Neimark-Sacker bifurcation at which two Floquet multipliers cross
from inside the unit circle, and the invariant curve of the mapping is born,
turns out to be a resonant phenomenon: independently of the values 
of $\mu$ and $B$ (the latter should be large enough to ensure
the existence of the fixed point), it occurs when the driving frequency equals
\begin{equation}
\label{neimark} 
\omega_{\rm NS}=
\frac{2\pi}{\tau_{\rm g}-\frac{\log\nu}{\lambda_+(1-\nu)}}\;\;\;
\end{equation}
For $\omega>\omega_{\rm NS}$, both periodic solutions, born in the
saddle-node bifurcation at $B=B_{\rm sn}$, are unstable. The place where
the Neimark-Sacker bifurcation branches off the saddle-node curve, 
lies on the right border of the Arnold tongue, at the highest point
of that border\footnote{The fixed point implies
$y=f(y)+B\,g(\Psi,\tau_{\rm loc}(y))$. 
Differentiation with respect to $\omega$ yields 
$$\frac{dB}{d\omega}=\frac
{\displaystyle 
1-f'(y)} {g(\Psi,\tau_{\rm loc}(y))}\frac{\partial y}{\partial\omega}
-\frac{\displaystyle (y-f(y))g'(\Psi,\tau_{\rm loc}(y))}
{g(\Psi,\tau_{\rm loc}(y))^2}\,
(\frac{\partial\Psi}{\partial\omega}
+\tau_{\rm loc}'(y)\frac{\partial y}{\partial\omega})$$
Since $g'=0$ along the saddle-node curve, the last term vanishes there. 
The Neimark-Sacker condition $f'(y)=1$ lets the first term vanish as well,
hence the codimension-two event occurs at the extremum of the saddle-node 
curve.}.
\begin{figure}[h]
\centerline{\includegraphics[width=0.7\textwidth]{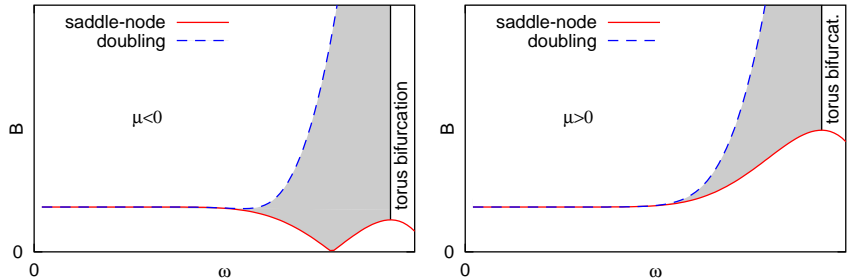}}
\caption{Bifurcation diagram for periodic $\varphi$-oscillations.
Gray regions: existence of stable periodic state. 
Left panel: $\mu<0$ (autonomous system is pre-homoclinic).
Right panel: $\mu>0$ (post-homoclinic situation).
}
\label{fig_6}
\end{figure}
Bifurcation diagrams for the fixed points of the mapping are shown
in Fig. \ref{fig_6}. Part of the diagram 
to the right from the Neimark-Sacker bifurcation is not relevant:
it refers to bifurcations of the distant fixed point of the autonomous 
mapping (\ref{combined_map}) with non-small value of $y$.

\subsubsection{Rotations}
For rotations, the Arnold tongue on the parameter plane
is bounded by the line
\begin{equation}
\label{sb_minus} 
B_{\rm sn}^{\rm r}(\omega)=|\,\mu-{\rm e}^{\lambda_+(\tau_{\rm g}-2\pi/\omega)}
                     +{\rm e}^{\nu\lambda_+(\tau_{\rm g}-2\pi/\omega)}\,|
\end{equation}

At positive values of $\mu$, 
when $\varphi$-rotations exist in the autonomous case as well,
the Arnold tongue opens from the abscissa $B=0$. 
For $\mu<0$, the region of existence of rotational solutions is
separated from the abscissa: without the force there are no rotations,
and in order to excite them, the frequency-dependent
threshold should be exceeded. \textit{Parts of the diagram to the
right from the curve of the torus bifurcation
refer to the unstable distant fixed
point, fictitious from the point of view of the underlying flow.}

Like in the case of oscillations, the region of stability of the
fixed point is bounded from above by the curve of the period-doubling
bifurcation which (again, like in that case) obeys Eq.(\ref{pd_osc})
so that at low frequencies the range of stability is nearly absent.
In contrast to the $\varphi$-oscillations, the period-doubling in the mapping
describes the genuine period-doubling of the underlying flow. 

The modulation instability of the periodic rotational solution occurs
at the same value of frequency (\ref{neimark}), as for the oscillations. 

Remarkably, at low values of the driving frequency the lower borders of
Arnold tongues for $\varphi$-oscillations and $\varphi$-rotations nearly 
coincide: the corresponding values of $B_{\rm sn}$ are separated by the
exponentially small term $2 {\rm e}^{\lambda_+(\tau_{\rm g}-2\pi/\omega)}$.
At negative $\mu$, the threshold for oscillations lies slightly lower; at
positive $\mu$, the opposite situation takes place. In its turn, this implies
that the borders of period-doubling bifurcation for rotations and oscillations
almost coincide as well.

\begin{figure}[h]
\centerline{\includegraphics[width=0.7\textwidth]{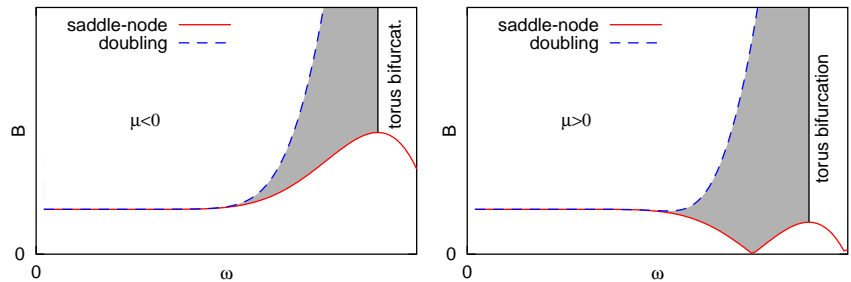}}
\caption{Bifurcation diagram for periodic $\varphi$-rotations.
Gray regions: existence of stable periodic state. 
Left panel: $\mu<0$ (autonomous system is pre-homoclinic).
Right panel: $\mu>0$ (post-homoclinic situation).}
\label{fig_7}
\end{figure}

The bifurcation diagram for periodic rotations is sketched 
in Fig.\ref{fig_7}. When the same values of $|\mu|$
are used for oscillations and rotations,
the panels of Fig.\ref{fig_7} look like
the interchanged panels of Fig.\ref{fig_6}.

Summarizing we see that in both cases, for oscillations and rotations, periodic
external modulation virtually does not induce phase locking 
at sufficiently low driving frequencies: 
the range of amplitude in which the locked periodic state is stable,
is exponentially small.

\section{Spin-torque oscillator under periodic force: bifurcations of
oscillatory states}
\label{sect_forced}
Let us apply results from the preceding section to the
case of the spin-torque oscillator influenced by the periodic force.
To this end, we
consider periodic modulation of the current with frequency 
${\Omega}$:  $I(t)=I_0+\eta \cos{\Omega} t$.

According to the general theory, self-sustained oscillations of the STO 
should be  entrained by the modulation, 
provided that $\Omega$
is sufficiently close to the frequency of the autonomous system; the larger
the mismatch between the frequencies, the higher should be the minimal
modulation amplitude which ensures entrainment.

Indeed, we observe entrainment for the values
of $I_0$ both below and above the gluing bifurcation in the autonomous
equations.
\begin{figure}[h]
\centerline{\includegraphics[width=0.9\textwidth]{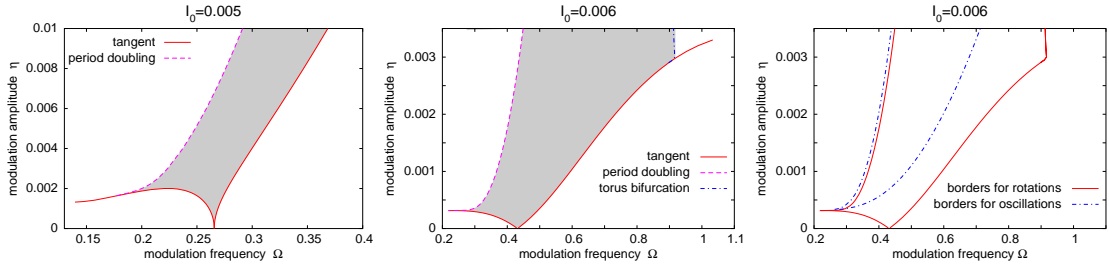}}
\caption{Main resonance regions on the parameter plane.
Values of $I_0$ correspond to dynamics below (left panel)
and above (center panel) the gluing bifurcation in the autonomous system.
Right panel: Stability regions for periodic $\varphi$-oscillations and 
$\varphi$-rotations on the parameter plane.}
\label{fig_8}
\end{figure}
In both cases, the 1$:$1 
entrainment occurs in the stripe of the parameter plane 
bounded by three curves: two curves of the tangent bifurcation 
and the curve of the period-doubling bifurcation,
beyond which the trajectory in the phase space closes after two periods
of modulation.

The shape of the resonant regions on the parameter plane bears
apparent distinctions from the familiar wedge-like form of the Arnold tongues:
the left borders are strongly deformed, the critical amplitude is a
non-monotonic function of the modulation frequency. As predicted by
analysis in the preceding section,  at low modulation frequencies
the interval of stability of the periodic state, 
is squeezed between the saddle-node bifurcation  
and the period-doubling bifurcation, and becomes extremely narrow.
In accordance with the prediction, the curve of the torus (Neimark-Sacker) 
bifurcation where the limit cycle yields to quasiperiodicity, is nearly
vertical: it can be encountered only in the rather narrow range of
modulation frequencies. Therefore, in Fig.~\ref{fig_8} this curve is shown
only for $I_0$=0.006; the values of $\Omega$ used for plotting
other bifurcation curves at $I_0$=0.005 in the left panel,  are too low.

Furthermore, in large parts of the resonance regions the driven system 
is multistable: depending on the initial conditions, trajectories can 
converge to either of two oscillatory states.
We demonstrate this effect in the right panel of Fig.~\ref{fig_8}, 
taking the value of $I$
above the threshold of the gluing bifurcation in the autonomous system. 
As seen in the plot, the main resonance region (bounded by red curves)
which corresponds to the rotation of $\varphi$, 
largely overlaps with the region (bounded by the blue curves) 
in which $\varphi$ performs localized oscillations.
The only difference is that at appropriate values of the modulation
frequency, $\varphi$-rotations can be excited by arbitrarily weak
driving, whereas the onset of   $\varphi$-oscillations (absent in the
autonomous system at this value of $I$) requires that the
modulation amplitude exceeds a certain finite threshold. 

\begin{figure}[h]
\centerline{ 
\includegraphics[height=40mm]{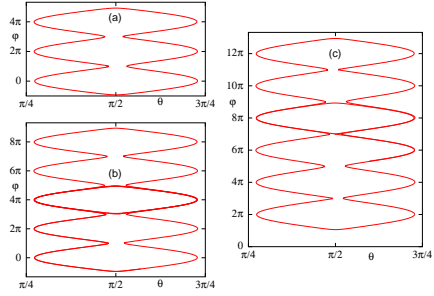}}
\caption{Projections of phase portraits for multi-turn periodic states.
$I_0$=0.006, $\Omega$=0.3. (a): $\eta$=0.002; (b): $\eta$=0.015; 
(c): $\eta$=0.0012.}
\label{fig_9}
\end{figure}

Besides the simple periodic attractors, in the large regions of
the parameter plane more complicated oscillations are observed: they include
multi-turn limit cycles, quasiperiodic  states as well as apparently
chaotic attractors. Detailed investigation of transitions between
those states lies outside the scope of the present paper; below we briefly
describe their main types.
Like simple limit cycles, the multi-turn ones
can be roughly divided into $\varphi$-oscillatory and $\varphi$-rotating 
ones; in Fig.~\ref{fig_9} we present a few characteristic phase portraits
for periodic states with zero net rotation along $\varphi$; during such
oscillations several cycles of clockwise rotations are followed by the same
amount of counterclockwise ones.

\begin{figure}[h]
\label{fig_10}
\centerline{ 
\includegraphics[height=40mm]{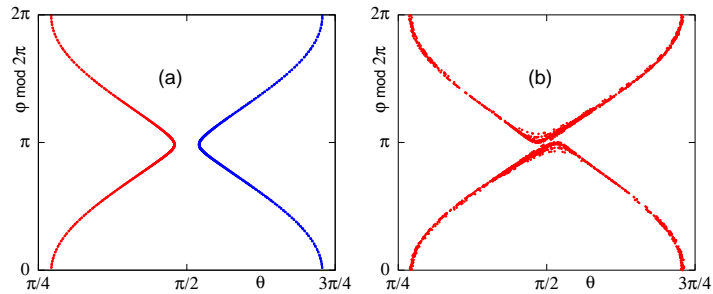}}
\caption{Stroboscopic mapping for quasiperiodic (a) and chaotic (b) 
oscillations. $I_0$=0.006, $\Omega$=0.3. 
(a): Two orbits at $\eta$=0.0005. Red dots: clockwise $\varphi$-rotation.
Blued dots: counterclockwise $\varphi$-rotation.
(b): $\eta$=0.0035. The single attracting orbit for which rotations 
in both directions alternate intermittently.}
\end{figure}

Quasiperiodic and weakly chaotic states are exemplified in Fig.~\ref{fig_10}
which shows the projections of stroboscopic mappings. 
Outside the Arnold tongues, at  low amplitudes
of the driving force the motions are quasiperiodic; the mapping possesses
two symmetric invariant curves which correspond to quasiperiodic rotations in
two opposite directions. When the amplitude is increased, the curves
approach each other and ``recombine''; coarsely, transition
to chaos occurs after this event. 

Altogether, as follows from theoretical predictions~\cite{Afr_Shil}, 
dynamics is quite rich, and the region where the STO just follows 
the external force occupies only a relatively 
small part of the parameter plane.

\section{Discussion}
\subsection{Summary of results}
We have characterized the typical features of entrainment by periodic
external force for systems close to a homoclinic (gluing) bifurcation.
General character of results allows to expect 
that similar deformation of Arnold tongues
with such consequences as multistability and/or chaos 
takes place in many periodically forced systems:
Unstable steady states play important dynamical role in a variety
of problems of fluid mechanics, nonlinear optics, physical chemistry, etc.,
and in many documented situations evolution of stable periodic regimes 
leads through homoclinic bifurcations of those states. 
In the above example the leading eigenvalues of the saddle are real.
Quite often the relevant steady states are of the saddle-focus type; 
dynamically, this brings in another oscillatory timescale 
which interacts with the existing ones, 
and makes the adequate description of entrainment much more involved.

We have considered the simple harmonic forcing; 
for a periodic force with many Fourier harmonics, the deformation of entrainment
region would be more drastic. This, for example, should be the case 
for a spin-torque
oscillator unilaterally influenced by another STO unit which,
itself, is close to homoclinicity. When the interaction is bilateral
(e.g. the STO are serially connected), robust synchronization of
two units can be restricted to small regions of the parameter space.

\subsection{Sketch of synchronization diagram for two coupled STO} 

\begin{figure}[h]
\centerline{\includegraphics[width=0.5\textwidth]{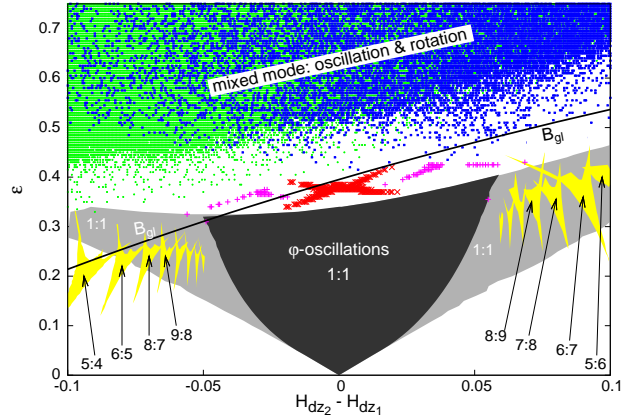}}
\caption{Sketch of the state diagram for two serially connected STO. 
Except for $\varepsilon$ and $H_{dz}$, the parameter values are taken 
from (\ref{parameters}). Current $I$ is fixed at 0.0055; $H_{dz_1}$=1.6; 
$\varepsilon$ and  $H_{dz_2}$ are varied.
White regions: quasiperiodic and chaotic asynchronous regimes with no
stable periodic states.
Black and gray regions: stable synchronous periodic $\varphi$-oscillations
of both units; in the black region they are globally attracting.
Yellow wedges: secondary resonance tongues with stable mode-locked states. 
Red crosses and magenta pluses: 
synchronous periodic states with, respectively, 2 and 4 turns in the 
phase space. Green and blue dots: coexisting mixed-mode states in which one
unit performs periodic $\varphi$-oscillations whereas the other one rotates. 
Curve $B_{\rm gl}$: gluing bifurcation for the isolated second STO.
}
\label{two_diagram}
\end{figure}

As an illustration, we take two identical or weakly non-identical serially 
connected STO; details of the setup are described e.g. in \cite{Pikovsky_2013}.
Each oscillator is characterized by the angles $\theta_i$ and $\varphi_i$
($i=1,2$) that, respectively, obey the first two equations of the system (\ref{eq_sto}).
The interaction occurs through the common current in the circuit,
affected by variations of magnetoresistance of both oscillators.
In the equations this is expressed in terms of the variable $u$ (rescaled voltage)
governed by 
$$\ddot{u}=\Omega_0^2\, 
       \bigg(\big( 1-\frac{\varepsilon}{2}
       (\sin\theta_1\cos\varphi_1+\sin\theta_2\cos\varphi_2)\big)
       \,(1-\frac{\dot{u}}{\omega})-u\bigg).$$
Thereby, intensity of interaction is regulated by the parameter $\varepsilon$.
In Fig. \ref{two_diagram} we sketch the state diagram 
obtained by variation of $\varepsilon$ and the difference between the
individual values of the parameter $H_{dz}$, responsible for the magnitude
of demagnetization field in the units.

Detailed analysis of bifurcations in the set of two coupled STO lies outside 
the scope of this paper and will be presented elsewhere. Here, we restrict 
ourselves to the most general features. The state diagram can hardly be 
directly compared to the treated above situation of periodic external driving, 
since the meaning of coordinates in Fig. \ref{two_diagram} 
and e.g.  in Fig.~\ref{fig_8}
is obviously different. This mostly concerns the ordinate of two diagrams.
In contrast to the modulation amplitude $\eta$ that becomes relevant 
only in presence of forcing, 
the parameter $\varepsilon$, 
responsible for the coupling, enters the governing equations of the uncoupled
STO oscillator as well: the corresponding term in the equation for $du/dt$
describes ``self-action'' of the oscillator caused by temporal dependence
of its magnetoresistance. Variation of $\varepsilon$, taken alone,
can result in bifurcations of Eqs (\ref{eq_sto}). As seen from Eq.(\ref{bif_Hopf}),
the Hopf bifurcation is $\varepsilon$-insensitive; in contrast, the gluing
bifurcation depend on $\varepsilon$. In the range $1.5<H_{dz}<1.7$, 
corresponding to the diagram of  Fig.~\ref{two_diagram}, 
an uncoupled STO performs
$\varphi$-oscillations at small values of $\varepsilon$, and $\varphi$-rotations
at the higher values of that parameter; transition between these 
two types of motion is mediated by the gluing bifurcation. Close to that
event, period of oscillations grows (recall Fig.~\ref{fig_4}b); thereby,
unlike the modulation amplitude $\eta$, the parameter $\varepsilon$
influences the timescale of an uncoupled unit. 
In the case of external driving, the parameter plane is divided by the
vertical line passing through the tip of the main resonance region: 
to the left from that line the uncoupled system is faster
than the driving force, and in the right half-plane the opposite situation 
takes place.
This clear dichotomy is absent in the case of two coupled units: at
fixed values of $H_{dz_1}$ and $H_{dz_2}$ there exist ranges 
of $\varepsilon$ in which the frequency of oscillations in the isolated first 
STO exceeds the frequency of the second one, as well as ranges of  
$\varepsilon$ where the second oscillator is faster. Therefore a mapping
between the regions of Fig. \ref{two_diagram} 
and Fig.~\ref{fig_8} is, to put it mildly, not straightforward.

At low values of $\varepsilon$, simple
synchronous $\varphi$-oscillations of both units have been observed in parts 
of the parameter plane (black and gray regions in Fig. \ref{two_diagram}).
However, they are globally attracting only at relatively small
values of $|H_{dz_1}-H_{dz_2}|$; otherwise, they coexist
with quasiperiodic or mode-locked motions, and their attraction basin 
shrinks when the difference between the values of $H_{dz}$ grows. This bistability
provides a distinction from the conventional picture of synchronization
in which quasiperiodic and periodic states do not coexist.
On the upper boundary of the region of synchronous $\varphi$-oscillations,
the Neimark-Sacker bifurcation takes place, and asynchronous quasiperiodic
regimes replace periodic oscillations. Quasiperiodicity, in turn, yields to chaos;
in certain regions of the plane we observe periodic windows where the
units perform multi-turn oscillations with zero net rotation of $\varphi$,
akin to the orbits from Fig.~\ref{fig_9}.

At higher values of $\varepsilon$ the periodic mixed-mode states are
present: one unit
performs $\varphi$-oscillations whereas the other one synchronously
rotates. In most of the cases, two such states coexist, albeit with
different size of attraction basins: in one of them
the unit with larger $H_{dz}$ oscillates and the other one rotates, 
in the second state the roles are interchanged.
Two parameter regions with synchronous periodic
attractors  at the bottom and at the top of the plot
are separated by the wide stripe with no stable periodic states. 
Remarkably, this stripe lies on both sides from the curve of the gluing bifurcation 
in the uncoupled unit. It should be noted that the latter curve does not
denote a bifurcation in the coupled system: there,
the saddle equilibrium has two-dimensional unstable manifold 
and cannot directly participate in bifurcations of stable periodic orbits. 
Nevertheless, in this region of the parameter plane the characteristic times 
of two oscillators lie sufficiently far apart, and sustainment of synchronous states
becomes hardly possible, both for identical ($H_{dz_1}$=$H_{dz_2}$) 
and non-identical STO oscillators. This confirms our conclusion that proximity
of separate units to the gluing bifurcation hampers synchronization.

\subsection{Relevance for experimental observations}
The last remark concerns a
 possible comparison of the above results to experimental observations.
It should be kept in mind, that in experiments at the nanoscales
it is difficult to separate purely deterministic features from noisy 
and often non-stationary background. The role of fluctuations
is not reduced to additive noise: e.g. the parameter drift can modulate 
intrinsic frequencies, and thereby cause and/or interrupt 
temporary epochs of synchrony. 
The appropriate theoretical description should include both the dynamical 
and the stochastic aspects.

\section*{Acknowledgement} Our research was supported by the Research Grant
PI 220/17-1 of the DFG.

\end{document}